\documentclass[submission,copyright,creativecommons]{eptcs}
\providecommand{\event}{FROM 2019} % Name of the event you are submitting to
\usepackage{underscore}           % Only needed if you use pdflatex.

\usepackage[cmex10]{amsmath}
\usepackage{cite}
\usepackage{mdwmath}
\usepackage{mdwtab}
\usepackage{verbatim}
\usepackage{amssymb}
\usepackage{doi}

\usepackage{amsthm}
\usepackage{textcomp}
\usepackage{color}
\usepackage{balance}
\newtheorem{property}{Property}
\newtheorem{definition}{Definition}
\newtheorem{myalgo}{Algorithm}

\newtheorem{conjecture}{Conjecture}

\newcommand{\altitem}[1]{\item {\em Alternative #1.}} % alternative item
\newcommand{\casitem}[1]{\item {\em Case #1.}} % case item
\newcommand{\caseq}[3]{\item {\em Case #1.}{\begin{equation}\label{#3}#2\end{equation}}} % case item with formula

\newcommand{\tma}{{\em Theorema }}

\renewcommand{\And}{\wedge}
\newcommand{\Aand}{{\ \ \wedge\ \ }}

\newcommand{\Implies}{\Longrightarrow}

\newcommand{\fa}[1]{{\underset{#1}{\forall}\ }}
\newcommand{\ex}[1]{{\underset{#1}{\exists}\ }}

\newcommand{\cons}{\smallsmile}
\newcommand{\leec}{\leq} % elem <= composite
\newcommand{\lec}{<} % elem < composite
\newcommand{\lece}{\leq}

\newcommand{\lecc}{\leq}
\newcommand{\mse}[1]{\{\!\!\{#1\}\!\!\}} % multiset with specified elements
\newcommand{\ms}{{\cal M}} % multiset
\newcommand{\union}{\uplus}
\newcommand{\el}{\langle \rangle} % empty list

\newcommand{\is}{\textit{IsSorted}}
\newcommand{\sort}{\textit{Sort}}

\newcommand{\conc}{\textit{Conc}}
\newcommand{\merge}{\textit{Merge}}

\newcommand{\trim}{\textit{Trim}}
\newcommand{\ins}{\textit{Insert}}
\newcommand{\mn}{\textit{min}}
\newcommand{\mnaux}{\textit{minA}}
\newcommand{\trimaux}{\textit{TrimA}}
\newcommand{\smalleq}{\textit{SmEq}}
\newcommand{\bigger}{\textit{Bigger}}

\newcommand{\rpr}[1]{{\bf Property \ref{#1}}} % ref property
\newcommand{\rdef}[1]{{\bf Definition \ref{#1}}} % ref definition
\newcommand{\rcj}[1]{{\bf Conjecture \ref{#1}}} % ref conjecture
\newcommand{\bpf}[2]{\noindent\begin{pf}\label{#2}{\em #1.}\\ } % begin proof
\newcounter{pf}
\newenvironment{pf}[1][]{\refstepcounter{pf}{\bf Proof \thepf: }{#1}}{{\em QED}}
\newcommand{\rpf}[1]{{\bf Proof \ref{#1}}} % ref proof

\newcommand{\bir}[2]{\noindent\begin{ir}\label{#2}{\em #1.} } % begin inference rule
\newcounter{ir}
\newenvironment{ir}[1][]{\refstepcounter{ir}{\bf IR-\their: }{#1}}{}
\newcommand{\rir}[1]{{\bf IR-\ref{#1}}} % ref inference rule

\newcommand{\bst}[2]{\noindent\begin{st}\label{#2}{\em #1.} } % begin strategy
\newcounter{st}
\newenvironment{st}[1][]{\refstepcounter{st}{\bf ST-\thest: }{#1}}{}
\newcommand{\rst}[1]{{\bf ST-\ref{#1}}} % ref strategy

\begin{document}
\title{Proof--Based Synthesis of Sorting Algorithms\\
  Using Multisets in \tma}
\def \titlerunning{Synthesis of Sorting Algorithms}

\author{Isabela Dr\u amnesc
\institute{Department of Computer Science\\
West University\\
Timi\c soara, Romania\\
Email: isabela.dramnesc@e-uvt.ro}
\and
Tudor Jebelean
\institute{Research Institute for Symbolic Computation,\\
Johannes Kepler University,\\
Linz, Austria\\
Email: Tudor.Jebelean@jku.at}
}
\def \authorrunning{I. Dr\u amnesc \& T. Jebelean}
\maketitle

\begin{abstract}
  Using multisets, we develop novel techniques for mechanizing the proofs of the synthesis conjectures for list--sorting algorithms, and we demonstrate them in the \tma system.
  We use the classical principle of extracting the algorithm as a set of rewrite rules
  based on the witnesses found in the proof of the synthesis conjecture produced from
  the specification of the desired function (input and output conditions).
  The proofs are in natural style, using standard rules, but most importantly
  domain specific inference rules and strategies.
  In particular the use of multisets allows us to develop powerful strategies for the synthesis of arbitrarily structured recursive algorithms by general Noetherian induction, as well as for the automatic generation of the specifications of all necessary auxiliary functions (insert, merge, split), whose synthesis is performed using the same method.
\end{abstract}

\section{Introduction}

We present a comprehensive case study in the automated synthesis of list sorting algorithms:
two main proofs produce the most popular sorting algorithms
(min--sort, quick--sort, insert--sort, merge--sort) and trigger all the proofs necessary
for producing the needed auxiliary functions for inserting, splitting, and merging.
This is a continuation of our work on exploring in parallel the theories of multisets,
lists, and binary trees, for the purpose of developing proof methods for the synthesis
of algorithms on these domains.
In one related paper \cite{sisy-2019} we already investigated algorithms for
deletion from lists and binary trees using multisets.

We follow the proof--based approach to automated synthesis:
first one proves automatically a {\em synthesis conjecture} which is based on the {\em specification} (input and output conditions) of the desired function, then the algorithm is extracted automatically from the proof, in form of conditional rewrite rules.
The theoretical basis and the correctness of this scheme is well--known  \cite{Bundy-deductiveproofs} and we used earlier in \cite{isa-jsc2015, isa-jsc2019}.

For the experiments we use the {\em Theorema} system \cite{theorema-2}, in which the logical formulae and the inferences are presented in {\em natural style}\footnote{That means a style similar to the one used by humans, but not {\em natural deduction}}, and which also allows to execute the synthesized algorithms.

\textbf{Related work.}
The theory of {\em multisets} is well studied in the literature, including
computational formalizations (see e. g. \cite{manna-waldinger1985-vol1}, where
finite multisets are called {\em bags}).
A presentation of the theory of multisets and a good survey of the literature related
to multisets and their usage is \cite{wayne-multiset-theory-1989} and some
interesting practical developments are in \cite{radoaca-multisets-and-sets-2015}.
A systematic formalization of the theory of lists using multisets for the proofs of correctness of various sorting algorithms is mechanized in Isabelle/HOL\footnote{
https://isabelle.in.tum.de/library/HOL/HOL-Library/Sorting_Algorithms.html}, which however does not address the problem of algorithm synthesis.
A valuable formalization in a previous version of \tma \cite{theorema-2000},
which includes the theory exploration and the synthesis of a sorting algorithm is
presented in \cite{Bruno-synthesis}, which also constituted the starting point of
our previous research on proof--based synthesis.
However, in that pioneering work, the starting point of the synthesis (besides the
specification of the desired function) is a specific {\em algorithm scheme}, while
in our approach we use general Noetherian induction and cover--set decomposition.
In our previous work we study proof--based algorithm synthesis in the theories of
lists \cite{saci-2012}, sets \cite{sisy-2014} and binary trees \cite{sisy-2015}
separately \cite{Synasc-2011}, \cite{Synasc-2012}, \cite{LATA2016}, \cite{isa-jsc2015},
\cite{isa-jsc2019}.

\textbf{Originality.}
In contrast to our early investigations, the current study
uses multisets, which leads to a crucial improvement of the proof techniques.
Also, the experiments are performed in the new version of the
\textit{Theorema} system \cite{theorema-2, theorema-2-windsteiger}.
More importantly, we do not use here algorithm schemata
or concrete induction principles, but only {\em general Noetherian induction} starting from
a specific {\em cover set} (usually based on the inductive definition of lists).
Namely, during the proof of a statement $P[t]$, for any $t'$  (also ground term) which
represents an object which is strictly smaller than the object represented by $t$
in the Noetherian ordering, $P[t']$ can be added to the current assumptions.
(The soundness of this technique is presented in detail in
\cite{isa-jsc2019} and it allows to discover concrete induction principles based on the
general Noetherian induction.)
In our approach we use the Noetherian ordering induced by the strict inclusion of the
corresponding multisets, which conveniently extends to a meta--ordering between terms,
induced by the strict inclusion of the constants occurring in the respective terms.

Moreover we develop a systematic approach to the {\em cascading} method pioneered in
\cite{Bruno-cascading}: when the
proof needs an auxiliary function which is not present in the knowledge,
the prover constructs a conjecture synthesis statement
which is used to obtain it.
We have been using cascading manually for the case of lists
in \cite{isa-jsc2015}, and in this paper we present it as automatic proof
technique and we illustrate it on several examples: all auxiliary
algorithms are generated by cascading starting from the sorting synthesis proofs.

For the purposes above, {\bf three novel inference rules} and {\bf six novel strategies} are introduced.

\section{Proof--Based Synthesis}\label{secII:proof-based-synthesis}

\subsection{Context}

{\bf Notation.}
Square brackets are used for function and for predicate application, for instance:
$f[x]$ instead of $f(x)$ and $P[a]$ instead of $P(a)$.
Quantified variables are placed under the quantifier, as in
$\underset{X }{\forall }$ and $\underset{X }{\exists }$.

\noindent{\bf Theory.}
We consider three types: {\em elements}, finite {\em lists}, and finite {\em multisets}.

\noindent{\bf Elements} (denoted by $a, b$) of lists are any objects whose domain is totally ordered (notation $\leq$ and $<$).
The ordering on elements is extended to orderings between an element and a list/multiset
and between lists/multisets, by requiring that all elements of the composite object observe the ordering relation\footnote{
  Note that this introduces exceptions to antisymmetry and transitivity when the empty
  list/multiset is involved.}.

\noindent{\bf Multisets} may contain the same elements several times.
$\emptyset$ denotes the empty multiset, $\mse{a}$ denotes the multiset having only the
element $a$ once.
The union (additive) is denoted by $\union:$ multiplicity is the sum of multiplicities --
like in \cite{Knuth-vol2}.
Union is commutative and associative with unit $\emptyset,$ these properties are used implicitly by the prover.
$\ms[U]$ denotes the multiset of elements of the list $U$.

\noindent{\bf Lists} (denoted $U, V, W$) are either empty $\el$ or of the form $a \cons U,$ where $\cons$ is the operation of prepending an element to a list (like {\em cons} of {\em Lisp}).
The multiset of a list observes:
\begin{property}\label{pr:list-multiset}
\mbox{ }
\(\underset{a,U}{\forall }\left(
\begin{array}{c}
\ms[\el]=\emptyset\\
\ms[a\cons U]=\mse{a} \union \ms[U]
\end{array}
\right)\)
\end{property}

\noindent
Sorted lists are defined by:
\begin{definition}\label{def:list-sorted}
\mbox{ }
\(\underset{a,U}{\forall }\left(
\begin{array}{c}
\is[\el]\\
\is[a\cons U] \Longleftrightarrow (a \leq U \wedge \is[U])
\end{array}
\right)\)\end{definition}

The {\em type of objects} is used by the prover, however for brevity we do not include the
type inferencing details in the proofs. In this presentation we just use an implicit
typing based on the notation convention.

\noindent
{\bf Problem and Approach.}
\label{sss:problem-approach}
The problem consists in finding the sorted version of a given list,
however by our approach several sub--problems may appear and require auxiliary algorithms
(merge, insert, split, etc.).
%Therefore we have to synthesize both functions with one argument (sort, split) and two arguments (insert, merge).
The synthesized algorithm is extracted from the proof of the
{\em synthesis conjecture} based on the function specification.
For univariate functions the specification consists in an
input condition $I[X]$ and an output condition $O[X, Y],$
and the conjecture is:
\begin{conjecture}\label{conj:univariate}
$\fa{X}(I[X]   \implies \ex{Y}O[X, Y]).$
\end{conjecture}

\noindent
Likewise, for a bivariate function one has $I[X, Y],$ $O[X, Y, Z],$ and the conjecture:
\begin{conjecture}\label{conj:bivariate}
$\fa{X,Y}(I[X,Y]   \implies \ex{Z}O[X, Y, Z]).$
\end{conjecture}

\subsection{Special Inference Rules and Strategies}\label{rules-strategy}

Following natural style proving, we use {\em Skolem constants} (denoted with numerical underscore like $V_1, a_0$) introduced for universal goals,
as well as {\em metavariables} (denoted with star power like $V^*,b^*_1$) introduced for existential goals.
The prover uses classical inference rules (split ground conjunctions, rewrite by equality, etc.)
as well as special rules appropriate for lists/multisets.
Some of these rules are already experimented in our previous work, and from those we list here only the ones which are used explicitely in the proofs presented in the paper.
The main contribution of this paper consists in the novel inference rules and strategies which construct the proofs necessary for the synthesis of sorting algorithms and their auxiliary functions, namely the inference rules:
\rir{ir:forward},
\rir{ir:backward}, and
\rir{ir:two-constants},
as well as the strategies:
\rst{st:cover-set},
\rst{st:induction},
\rst{st:cascading},
\rst{st:pair-MS},
\rst{st:split}, and
\rst{st:split-goal-eq}.

\subsubsection{Inference Rules}
\mbox{ }

% IR-
\bir{Forward inference}{ir:forward}% IR-
If a ground atomic assumption matches a part of another (typically universal) assumption,
instantiate the later and replace in it the resulting copy of the ground assumption by the constant
{\em True}, then simplify truth constants to produce a new assumption.
% This rule is akin to unit resolution and realizes for instance Modus Ponens and Modus Tollens.
It is used for instance in proving the goal (\ref{goal:list-sort-7}) (after the instantiation with the witnesses) on the basis of assumption (\ref{assm:list-sort-3}).
\end{ir}

% IR-
\bir{Backward inference}{ir:backward}% IR-
Transform the goal using some assumption or a specific logical principle.
If a ground atomic assumption matches a part of a ground or existential goal,
instantiate the later and replace in it the resulting copy of the ground assumption by the constant
{\em True}, then simplify truth constants to produce a new goal.
%(The rule \rir{ir:assumed-subgoal} is a special case of this when there is no variable involved.)
% not used explicitely in the displayed proofs
A specific logical principle is used for backward inference on goals containing metavariables,
namely the fact that a formula having the structure $\ex{x}P[x]$ is a logical consequence of
the formula $\ex{x}P[f[x]].$ Example: transformation of (\ref{goal:list-sort-4a}) into (\ref{goal:list-sort-4}).
\end{ir}

% IR-
\bir{Reduce composite argument}{ir:red-comp-arg}% IR-
This rule uses the current knowledge to transform parts of the goal or of the assumptions into atoms
whose arguments contain no function symbols.
Example: (\ref{goal:list-sort-3}) and (\ref{goal:list-sort-4a}).
%\noindent Example: $a_0 \leec \conc[U_0, Y_0]$ becomes $a_0 \leec U_0 \And a_0 \leec Y_0.$
\end{ir}

% IR-
\bir{Solve metavariable}{ir:solve-meta}% IR-
When the goal is $\ms[X^*] = \ms[{\cal T}]$ for a ground term ${\cal T},$ infer
$X^* = {\cal T}.$
Example: formula (\ref{goal:sort-list-13}).
Sometimes this involves several intermediate steps -- see (\ref{goal:list-sort-4a}) -- (\ref{assm:list-sort-3}).
\end{ir}

% IR-
\bir{Expand multiset}{ir:expand-multiset}% IR-
In the goal, a multiset term with a composite argument is expanded by equality into several
multiset terms.
This is typically used when the argument contains cover--set constants, because about these we
do not have much information in the assumptions, but by treating them separately we can obtain
objects having more properties, for instance by applying induction.
Example: (\ref{eq:sort-list-goal-0}) -- (\ref{goal:sort-list-11}).
\end{ir}

% IR-
\bir{Compress multiset}{ir:compress-multiset}% IR-
This is the dual of the previous rule, and it is tipically applied when the arguments contain function calls introduced by induction or by cascading.
Example: (\ref{goal:sort-list-22}) -- (\ref{goal:sort-list-23}).
\end{ir}

% IR-
\bir{Use equivalence}{ir:equivalence} % IR-
Equality of the corresponding multisets induces an equivalence relation on lists,
which is compatible with the ordering relations induced by the domain ordering, as well as
with the function $\sort.$
Therefore the prover can rewrite parts of the goal or of the assumptions by replacing equivalent lists or by inferring new relations on lists which are equivalent to lists already related.
Example: (\ref{goal:list-sort-5}) -- (\ref{goal:list-sort-7}).
\end{ir}

% IR-
\bir{Two constants}{ir:two-constants}% IR-
If the current proof situation contains two Skolem constants representing domain elements,
say $a_0, b_0$, then the prover generates two cases: $a_0 \leq b_0$ and $b_0 < a_0.$
\noindent
Example: after (\ref{goal:min-aux-trim-aux-3}).
\end{ir}

\subsubsection{Strategies}

\mbox{ }

% ST-
\bst{Cover set}{st:cover-set}
  This strategy organizes the structure of each synthesis conjecture proof
  and the extraction of the synthesized algorithm.
  Each conjecture for the synthesis of a {\em target function} is a quantified statement
  over some {\em main universal variable}. % which is the initial goal of the synthesis proof
  A {\em cover set} is a set of universal terms\footnote{
  Terms containing universally quantified variables, such that for every element of the domain
  there exists exactly one term in the set which instantiates to that element.}
  which represent the domain of the main universal variable, as described in \cite{isa-jsc2019}.

  We project this concept on Skolem constants: first the main universal variable is
  Skolemized (``arbitrary but fixed'') --- we call this the {\em target constant},
  and we call the corresponding Skolemized goal the {\em target goal} --
  and then the corresponding cover--set terms are also grounded by Skolemization,
  we call these the {\em cover-set terms} and the corresponding constants the
  {\em cover-set constants}.
  The proof starts with a certain cover set (typically the one suggested
  by the recursive definition of the domain), and starts a proof branch for each ground term
  (``proof by cases'') -- see {\em Alternative 2} in \rpf{pf:sort-list-def}.
  On each proof branch the input conditions of the function are assumed, and then the
  existential variable corresponding to the output value of the function is
  transformed into a metavariable whose value (the ``witness'') will be found
  on the respective branch of the proof.
  Finally the algorithm will be generated as a set of [conditional] equalities:
  the terms of the cover set become arguments (``patterns'') on the LHS of the
  equalities, and the corresponding witnesses become the RHS of these, after replacing
  back the Skolem constants by variables.
  The strategy can be applied in a {\em nested} way, by choosing a new target constant among the Skolem constants of the goal -- see {\em Alternative 2.3} in \rpf{pf:merge}.

  The strategy is applied similarly to a metavariable from the goal (see {\em Alternative 1} in \rpf{pf:sort-list-def}),
  here the variables of the cover--set terms are replaced by metavariables.
  If on some branch the cover--set term is constant (it contains no metavariables), then the
  solution is constant and it may impose certain conditions on the Skolem constants
  involved in the goal, which  will be used as conditions on the inputs (which correspond to
  the respective Skolem constants) in the final expression of the algorithm.
  In order to ensure mutual exclusion, the negation of these conditions are
  transmitted as additional assumptions to the next branches -- see formula (\ref{assm:list-sort-1}).
\end{st}

% ST-3
\bst{Induction}{st:induction}
  We use Noetherian induction based on the well--founded ordering between lists
  determined by the strict inclusion of the corresponding multisets.
  This ordering checked either syntactically by the meta-relation between terms induced by the
  strict inclusion of the multisets of constants occurring in the terms,
  either semantically by using the current assumptions:
  for instance if $a_0\cons U_0$ is a cover--set term for the target constant $X_0$
  then $U_0$ is smaller than $X_0.$

  When a ground term $t$ represents an object which is smaller than the target constant $X_0$
  of the target goal $P[X_0]$, then $P[t]$ is added as a new assumption, but modified by
  inserting the corresponding call of the target function instead of the existential variable.

  Example: the target function is $F[X,Y],$ the target constant is $X_0,$
  the target goal $P[X_0]$ is $\fa{Y}(I[X_0,Y] \implies \ex{Z}O[X_0,Y,Z]),$
  and we have a ground term $t$ smaller than $X_0$ in the well--founded ordering.
  The instance $P[t]$ of the target goal is $\fa{Y}(I[t,Y] \implies \ex{Z}O[t,Y,Z]),$
  The prover adds the assumption $\fa{Y}(I[t,Y] \implies O[t,Y,F[t,Y]]),$
  Typically in the subsequent proof this will be instantiated with a ground term $s,$
  then $I[t,s]$ will be proven and $O[t,s,F[T,s]]$ will be obtained as assumption,
  leading to the replacement of some subterm[s] of the goal with $F[t,s].$
  In this way the recursive calls of $F$ are explicitly generated in the synthesized algorithm,
  see for instance formulae (\ref{assm:merge-a10.1}) to (\ref{goal:merge-g11}).

  This strategy is applied in a similar manner to metavariables, when they occur in the goal.
  When a metavariable $Y^*$ represents an object which is smaller than the target constant $X_0,$
  then $P[Y^*]$ may be added as new assumption -- see formula (\ref{assm:list-sort-1a}).
\end{st}

% ST-4
\bst{Cascading}{st:cascading}
  This strategy consists in proving separately a conjecture for synthesizing the algorithm for some
  auxiliary functions needed in the current proof.
  The Skolem constants from the current goal become universal variables $x, x', \dots,$
  the metavariables from the current goal become existential variables $y, y', \dots,$
  and the conjecture has the structure\footnote{By local convention, here $x, x', y, y'$ represent any
  kind of objects: domain elements or lists.}:
  \begin{equation}\label{eq:cascade-general}
  \fa{x}\fa{x'}\ldots(P[x, x', \ldots] \Implies \ex{y}\ex{y'}\ldots Q[x, x', \ldots, y, y', \ldots])
  \end{equation}
  $P[x, x', \ldots]$ is composed from the assumptions which contain {\em only} the Skolem constants
  present in the goal, and $Q[x, x', \ldots, y, y', \ldots]$ is composed from the goal.
  A successfull proof of the conjecture generates the functions
  $f[x, x', \ldots], f'[x, x', \ldots], \ldots,$ which have the property:
  \begin{equation}\label{eq:cascade-property}
  \fa{x}\fa{x'}\ldots(P[x, x', \ldots] \Implies
                           Q[x, x', \ldots, f[x, x', \ldots], f'[x, x', \ldots], \ldots])
  \end{equation}
  The current proof continues after adding this property to the assumptions, thus if some of the
  generated functions are necessary later in the proof, they can be used without a new cascading step.
  Similar to the situation described at \rst{st:induction}, the new assumption will trigger the simplification of the current goal by inserting the auxiliary function -- see for instance formulae (\ref{assm:sort-list-14}) and (\ref{goal:sort-list-13}).
\end{st}

% ST-5
\bst{Pair multisets}{st:pair-MS}
  This strategy applies when the goal contains an equality of the shape:
  $\ms[Y^*] = \ms[t_1] \union  \ms[t_2] \union \ldots,$
  where $Y^*$ is the metavariable we need to solve, and $t_1, t_2, \ldots$ are ground terms.
  A typical flow of the proof consists in transforming the union on the RHS of the equality
  into a single $\ms[t]$, because this gives the solution $Y^* \rightarrow t.$
  To this effect the prover groups pairs of operands of $\union$ together
  (no matter whether they are contingent or not, because commutativity),
  creating alternatives for different groupings.
  For each pair a conjecture is created as described at strategy \rst{st:cascading} (cascading),
 from which a multiset term which equals the union of the pair can be constructed in one of
  the following ways:
  \begin{itemize}
  \item \hspace{-15pt}-- the auxiliary function is already known, the proof works by predicate logic;% -- see \ralg{algorithm:sorted-lists-merge-4};
  \item \hspace{-15pt}-- induction can be applied (if the target function is binary) - see formula (\ref{ec:concat-lists-sorted-g2-cascade-v2});
  \item \hspace{-15pt}-- a separate synthesis proof of the function is necessary by \rst{st:cascading} (cascading) -- see \rcj{conj:insert}.
  \end{itemize}
\end{st}

%ST-6
\bst{Split}{st:split}
 When a union of multisets in the RHS of the goal must be sorted
 and it contains $\mse{a}$ and $\ms[X]$ where $a$ and $X$ are incomparable,
 split $X$ into $X_1, X_2$ such that $X_1 \leq a$ and $a < X_2.$
 Similarly to the situation shown at \rst{st:pair-MS}, the two lists are found either by already known auxiliary functions, by induction, or by cascading,
 and the goal is updated appropriately with the corresponding terms.
 Example: {\em Alternative 2.2.2} in \rpf{pf:sort-list-def}.
\end{st}

% ST-8
\bst{Split goal equation}{st:split-goal-eq}
When the goal contains several metavariables in an equation, then split the equation into several
ones, such that only one metavariable occurs in every new equation.
Uses heuristics to match the appropriate values.
Example: formulae (\ref{goal:min-aux-trim-aux-4}) and (\ref{goal:min-aux-trim-aux-4a}).
\end{st}

\section{Synthesis of Sorting}

The experiments start with the synthesis of {\em sorting} --- the target function is $\sort$.
By cascading this will trigger the synthesis of other auxiliary algorithms for
insertion, merging, and splitting.
According to \rcj{conj:univariate} the synthesis conjecture is:
\begin{conjecture}\label{conj:sort-list}
$\fa{X}\ex{V}(\ms[V] = \ms[X]\And \is[V]).$
\end{conjecture}

\bpf{Sort list by definition--based cover set}{pf:sort-list-def}
  Universal $X$ is Skolemized to target constant $X_0,$ producing the target goal:
  \begin{equation}\label{goal:list-sort-target}
    \ex{V}\ms[V] = \ms[X_0]\Aand \is[V]
  \end{equation}
  and the existential $V$ becomes the metavariable $V^*$:
  \begin{equation}\label{goal:list-sort-0}
    \ms[V^*] = \ms[X_0]\Aand \is[V^*].
  \end{equation}
  Two alternatives are pursued, by applying strategy \rst{st:cover-set} (cover set) to the
  metavariable $V^*$ or to the Skolem constant $X_0:$
  %\begin{itemize} % alternatives 0

    \noindent%\altitem{1}
    {\em Alternative 1:}
    Apply \rst{st:cover-set} (cover set) to $V^*$ with the cover set determined by the domain
    definition: $\{\el,\ a^*\cons U^*\}$
    \begin{itemize} % cases 1
      \casitem{1.1} $V^*=\el$:
      The goal (\ref{goal:list-sort-0}) becomes:
      \begin{equation}\label{goal:list-sort-1}
        \ms[\el] = \ms[X_0]\Aand \is[\el].
      \end{equation}
      By inference rule \rir{ir:backward} (backward inference) using \rdef{def:list-sorted} the goal
      (\ref{goal:list-sort-1}) becomes:
      \begin{equation}\label{goal:list-sort-2}
        \ms[\el] = \ms[X_0].
      \end{equation}
      By \rst{st:cover-set} (cover set) the proof succeeds on this branch,
      the witness is $\el,$ the condition on the input is $X = \el,$ and the cumulated
      condition on the input for the next branch is $X_0 \neq \el.$

      \casitem{1.2} $V^*=a^*\cons U^*$:
      The condition on $X_0$ from the previous branch is added as assumption:
      \begin{equation}\label{assm:list-sort-1}
        X_0 \neq \el.
      \end{equation}
      The goal (\ref{goal:list-sort-0}) becomes:
      \begin{equation}\label{goal:list-sort-3}
        \ms[a^*\cons U^*] = \ms[X_0]\Aand \is[a^*\cons U^*]
      \end{equation}
      and the current solution for $V^*$ is $a^*\cons U^*.$
      By inference rule \rir{ir:red-comp-arg} (reduce composite argument)
      using \rdef{def:list-sorted} the goal (\ref{goal:list-sort-3}) becomes:
      \begin{equation}\label{goal:list-sort-4a}
        \ms[a^*\cons U^*] = \ms[X_0]\Aand a^*\leec U^* \Aand \is[U^*].
      \end{equation}
      By \rir{ir:backward} (backward inference) $U^*$ is replaced by $\sort[W^*]$ and
      the goal becomes:
      \begin{equation}\label{goal:list-sort-4}
        \ms[a^*\cons \sort[W^*]] = \ms[X_0]\Aand a^*\leec \sort[W^*] \Aand \is[\sort[W^*]]
      \end{equation}
      and the intermediate solution for $V^*$ is $a^*\cons \sort[W^*].$
      Since $a^*\cons \sort[W^*]$ stands for $V^*$ which has the same elements as the target
      constant $X_0$, the prover infers that $W^*$ is less than $X_0$ in the well--founded
      ordering, thus by strategy \rst{st:induction} (induction) the target goal (\ref{goal:list-sort-target}) is used with
      $\{X \rightarrow W^*\}$ and $\{V \rightarrow \sort[W^*]\}$ to generate the assumption:
      \begin{equation}\label{assm:list-sort-1a}
        \ms[\sort[W^*]] = \ms[W^*]\Aand \is[\sort[W^*]].
      \end{equation}
      The second conjunct of this assumption is used to reduce the goal (\ref{goal:list-sort-4})
      by rule \rir{ir:backward} to:
      \begin{equation}\label{goal:list-sort-5}
        \ms[a^*\cons \sort[W^*]] = \ms[X_0]\Aand a^*\leec \sort[W^*].
      \end{equation}
      The first conjunct is used by \rir{ir:equivalence} (use equivalence) to reduce the last goal to:
      \begin{equation}\label{goal:list-sort-7}
        \ms[a^*\cons W^*] = \ms[X_0]\Aand a^*\leec W^*.
      \end{equation}
      The strategy \rst{st:cascading} (cascading) is applied to this goal and generates the conjecture:
      \begin{conjecture}\label{conj:min-Trim}% was conj:list-sort-8
        $\fa{X}(X \neq \el \Implies \ex{a}\ex{U}(\ms[a\cons U] = \ms[X]\Aand a\leec U)).$
      \end{conjecture}
      \rpf{pf:list-min-trim} synthesizes the functions
      $\textit{min}[X]$ and $\textit{Trim}[X]$ which split a list into its minimum and the rest.
      By \rst{st:cascading} (cascading) the new assumption is:
      \begin{equation}\label{assm:list-sort-3}
        \fa{X}(X \neq \el \Implies
        (\ms[\mn[X]\cons\trim[X]] = \ms[X]\Aand
        \mn[X]\leec \trim[X])).
      \end{equation}
      Using (\ref{assm:list-sort-1}) this solves the goal (\ref{goal:list-sort-7}) with
      the witnesses: $\{a^*\rightarrow \mn[X_0]\}$ and $\{W^*\rightarrow\trim[X_0]\} ,$
      which gives for $V^*$ the final solution $\mn[X_0]\cons \sort[\trim[X_0]].$
      The algorithm extracted from the proof is:
      \begin{myalgo}\label{algorithm:min-sort} {\em Min-Sort.}
        \mbox{ }\\
        \(\fa{U}\left(
        \begin{array}{c}
          \sort[\el]=\el\\
          U\neq\el\Implies\sort[U]= \mn[U]\cons\sort[\trim[U]]
        \end{array}
        \right)\)
      \end{myalgo}
    \end{itemize} % cases

    %\altitem{2}
    \noindent
    {\em Alternative 2:}
    Apply \rst{st:cover-set} on $X_0$ with the cover set $\{\el,\ a_0\cons U_0\},$
    starting two branches:
    \begin{itemize}
      \casitem{2.1} $X_0 = \el$ is straightforward. The solution is $\{V^* \rightarrow \el\}.$
      \casitem{2.2} $X_0 = a_0\cons U_0:$
      The goal becomes:
      \begin{equation}\label{eq:sort-list-goal-0}
        \ms[V^*] = \ms[a_0\cons U_0]\And \is[V^*].
      \end{equation}
      By \rir{ir:expand-multiset} (expand multiset) using \rpr{pr:list-multiset}
      the goal is transformed into:
      \begin{equation}\label{goal:sort-list-11}
        \ms[V^*] = \mse{a_0}\union\ms[U_0]\And \is[V^*].
      \end{equation}
      Two alternatives are pursued, depending on the strategy used for this goal (\rst{st:induction} or  \rst{st:split}).
      \begin{itemize} % alternatives 1
        \altitem{2.2.1}
        Strategy \rst{st:induction} (induction) uses $U_0$ (smaller than $X_0$) to produce the
        assumption:
        \begin{equation}\label{assm:sort-list-11}
          \ms[\sort[U_0]] = \ms[U_0]\ \ \And\ \ \is[\sort[U_0]].
        \end{equation}
        The goal (\ref{goal:sort-list-11}) is rewritten by equality (\ref{assm:sort-list-11}) into:
        \begin{equation}\label{goal:sort-list-12}
          \ms[V^*] = \mse{a_0}\union\ms[\sort[U_0]]\Aand \is[V^*].
        \end{equation}
        Strategy \rst{st:pair-MS} (pair multisets) applied to $\mse{a}$ and $\ms[\sort[U_0]]$, using (\ref{assm:sort-list-11}) and (\ref{goal:sort-list-12}),
        produces the conjecture:
        \begin{conjecture}\label{conj:insert} %was conj:sort-list-10
          $\fa{a}\fa{X}(\is[X]\implies\ex{V}(\ms[V] = \mse{a}\union\ms[X]\And \is[V])).$
        \end{conjecture}
        The function $\ins[a,X]$ which inserts an element in a sorted list, keeping it sorted, is synthesized by the same method\footnote{For space reasons the proof is not included in this paper.}.
        By strategy \rst{st:cascading} (cascading) the new assumption is:
        \begin{equation}\label{assm:sort-list-14}
          \fa{a}\fa{X}(\is[X]\implies(\ms[\ins[a,X]] = \mse{a}\union\ms[X]\And \is[\ins[a,X]]))
        \end{equation}
        and the goal (\ref{goal:sort-list-12}) becomes:
        \begin{equation}\label{goal:sort-list-13}
          \ms[V^*] = \ms[\ins[a_0,\sort[U_0]]]\And \is[V^*].
        \end{equation}
        By \rir{ir:solve-meta} (solve metavariable) the solution for
        $V^*$ is $\ins[a_0,\sort[U_0]]$
        and the proof succeeds by standard logical inferences, thus the algorithm is:
        \begin{myalgo}\label{algorithm:insert-sort} {\em Insert-Sort.}
          \mbox{ }\\
          \(\fa{a,U}\left(
          \begin{array}{c}
            \sort[\el]=\el\\
            \sort[a\cons U]= \ins[a,\sort[U]]
          \end{array}
          \right)\)
        \end{myalgo}

        \altitem{2.2.2}
        The RHS of the equality in the goal (\ref{goal:sort-list-11}) represents
        a list which must be sorted and it contains $\mse{a_0}$ and $\ms[U_0],$
        where $a_0$ and $U_0$ are incomparable by the current assumptions.
        Therefore the strategy \rst{st:split} (split) applies to generate the conjecture:
        \begin{conjecture}\label{conj:lesseq-bigger} %was {conj:sort-list-20}
          $\fa{a}\fa{X}\ex{V_1}\ex{V_2}(\ms[X] = \ms[V_1]\union\ms[V_2]\Aand
          V_1 \leq a \Aand a < V_2).$
        \end{conjecture}
        \rpf{pf:split} of this conjecture generates the algorithms for the functions $\smalleq[a,X]$ and\\ 
        $\bigger[a,X]$ which split the list $X$ into two lists having
        elements which are smaller, respectively bigger than $a.$

        By strategy \rst{st:cascading} (cascading) the new assumption is:
        \begin{equation}\label{assm:sort-list-20}
          \fa{a}\fa{X}(\ms[X] = \ms[\smalleq[a,X]]\union\ms[\bigger[a,X]]
          \Aand\smalleq[a,X] \leq a \Aand a < \bigger[a,X]).
        \end{equation}
        By strategy \rst{st:split} (split) this is instantiated with $a_0$ and $U_0$ to produce:
        \begin{equation}\label{assm:sort-list-21}
        \begin{split}
          \ms[U_0] = \ms[\smalleq[a_0,U_0]]\union\ms[\bigger[a_0,U_0]]\ \ \And \\
          \smalleq[a_0,U_0] \leq a_0\ \  \And\ \  a_0 < \bigger[a_0,U_0]).
          \end{split}
        \end{equation}
        and the goal (\ref{goal:sort-list-11}) is transformed into:
        \begin{equation}\label{goal:sort-list-21}
          \ms[V^*] = \ms[\smalleq[a_0,U_0]]\union\mse{a_0}\union\ms[\bigger[a_0,U_0]]
          \Aand \is[V^*].
        \end{equation}
        Because (\ref{assm:sort-list-21}) neither of $\smalleq[a_0,U_0]$ and $\bigger[a_0,U_0]$
        can have more elements than $U_0$ and this is smaller in the well--founded ordering
        than the target constant $X_0$ because it is a part of a cover--set term.
        Thus strategy \rst{st:induction} (induction) is applied to both, producing assumptions:
        \begin{equation}\label{assm:sort-list-25}
          \ms[\smalleq[a_0,U_0]] = \ms[\sort[\smalleq[a_0,U_0]]]\ \ \And\ \
          \is[\sort[\smalleq[a_0,U_0]]],
        \end{equation}
        \begin{equation}\label{assm:sort-list-27}
          \ms[\bigger[a_0,U_0]] = \ms[\sort[\bigger[a_0,U_0]]]\ \  \And\ \
          \is[\sort[\bigger[a_0,U_0]]].
        \end{equation}
        Rewriting using (\ref{assm:sort-list-25}) and (\ref{assm:sort-list-27})
        replaces in the goal (\ref{goal:sort-list-11}) the corresponding subterms to obtain:
        %\vspace{-10pt}
        \begin{equation}\label{goal:sort-list-22}
          \ms[V^*] = \ms[\sort[\smalleq[a_0,U_0]]]\union\mse{a_0}
          \union\ms[\sort[\bigger[a_0,U_0]]]\And \is[V^*].
        \end{equation}
        Using \rir{ir:compress-multiset} by \rpr{pr:list-multiset} this becomes:
        \begin{equation}\label{goal:sort-list-23}
          \ms[V^*] = \ms[\sort[\smalleq[a_0,U_0]]]\union
          \ms[a_0 \cons \sort[\bigger[a_0,U_0]]]\Aand \is[V^*].
        \end{equation}
        By \rir{ir:forward} (forward inference) using the current assumptions and the properties of inequality the following are obtained:
        $\sort[\smalleq[a_0,U_0]] \lece a_0 \lec \sort[\bigger[a_0,U_0]]],$
          $\is[a_0 \cons \sort[\bigger[a_0,U_0]]]$ and
          $\sort[\smalleq[a_0,U_0]] \lecc a_0 \cons \sort[\bigger[a_0,U_0]].$

        Strategy \rst{st:pair-MS} applied to $\ms[\sort[\smalleq[a_0,U_0]]]$ and $\ms[a_0 \cons \sort[\bigger[a_0,U_0]]]$
        produces:
        \begin{conjecture}\label{conj:concat} %was {goal:split-list-0a}
          $\fa{X}\fa{Y}((X \leq Y \And \is[X] \And \is[Y]) \Implies
          \ex{V}(\ms[V] = \ms[X]\union\ms[Y] \And
          \is[V])).$
        \end{conjecture}
        The algorithm $\conc$ which concatenates two lists into a sorted one, if the conditions are like above, is also synthesized by our prover\footnote{For lack of space the proof is not presented in this paper}.
        The new goal is:
        \begin{equation}\label{goal:sort-list-23a}
          \ms[V^*] = \ms[\conc[\sort[\smalleq[a_0,U_0]]],
            a_0 \cons \sort[\bigger[a_0,U_0]]]\Aand \is[V^*],
        \end{equation}
        which gives the obvious solution to $V^*$ and the algorithm {\em Quick-Sort}:
        \begin{myalgo}\label{algorithm:quick-sort} {\em Quick-Sort.}
          \mbox{ }\\
          \(\fa{a,U}\left(
          \begin{array}{c}
            \sort[\el]=\el\\
            \sort[a\cons U]= \conc[\sort[\smalleq[a,U]], a \cons \sort[\bigger[a,U]]]
          \end{array}
          \right)\)
        \end{myalgo}
      %\end{itemize} % alternatives 1
    \end{itemize} % cases by cover set
  \end{itemize} % alternatives 0
\end{pf}

\smallskip

Another approach is to consider a cover set corresponding to the {\em divide--and--conquer}
principle:
  $\{\el, \ \ a\cons\el,\ \ \conc[U, V]\}$
(where $U, V$ are nonempty).
Here $\conc$ is used as a {\em pattern matching} construct, which may appear on the
LHS of a rewrite rule, and it comes together with a simple splitting function,
which gives two nonempty lists from a list having at least two elements.
(For lack of space we omit here a possible splitting algorithm and its automatic
generation by the principles presented in this paper.)
The proof proceeds in a similar manner, with several alternatives and successful branches,
from which we summarize below only the most interesting ones.

\bpf{Sort list by divide--and--conquer cover set}{pf:sort-list-dac}
By quantified inferences the target goal is the same as in the previous proof:
  \begin{equation}\label{goal:list-sort-dac-0}
    \ms[V^*] = \ms[X_0]\Aand \is[V^*].
  \end{equation}
%\begin{itemize} % alternatives 0
  %\altitem{1}

\noindent
{\em Alternative 1:}
  Application of the cover--set strategy to metavariable $V^*$ produces {\em Quick--Sort.}

  %\altitem{2}
\noindent
{\em Alternative 2:}
  Application of the cover--set strategy to $X_0.$
  Cases $\el$ and $a_0\cons\el$ are straightforward.
  %\begin{itemize} % cases
    %\casitem{1} $\el$ straightforward.

    %\casitem{2} $a_0\cons\el$ straightforward.

    %\casitem{3}

     \noindent
     { Case} $X_0 = \conc[U_1,U_2]:$
       After splitting the multiset the goal becomes:
       \begin{equation}\label{goal:list-sort-dac-1}
         \ms[V^*] = \ms[U_1]\union\ms[U_2]\Aand \is[V^*].
       \end{equation}
       After applying \rst{st:induction} (induction)\footnote{Note that induction can be applied only when $U_1, U_2$ are assumed
         nonempty.} on $U_1$ and on $U_2$ (we do not list the obvious assumptions):
       \begin{equation}\label{goal:list-sort-dac-2}
         \ms[V^*] = \ms[\sort[U_1]]\union\ms[\sort[U_2]]\Aand \is[V^*].
       \end{equation}
       Strategy \rst{st:pair-MS} (pair multisets) produces the conjecture:
       \begin{conjecture}\label{conj:merge}
       \mbox{ }\\
         $\fa{U_1,U_2}(\is[U_1]\And\is[U_2]\Implies\ex{W}(\ms[W] = \ms[U_1]\union\ms[U_2]\Aand \is[W])).$
       \end{conjecture}
       The proofs in section \ref{sec:merge} synthesize several algorithms for the function $\merge$ which combines two sorted lists into a sorted one.
       The corresponding sorting algorithm is:
        \begin{myalgo}\label{algorithm:merge-sort} {\em Merge Sort.}
          \mbox{ }\\
          \(\fa{a,U,V}\left(
          \begin{array}{c}
            \sort[\el]=\el\\
            \sort[a\cons\el] = a\cons\el\\
            \sort[\conc[U,V]] = \merge[\sort[U], \sort[V]]
          \end{array}
          \right)\)
        \end{myalgo}
  %\end{itemize} % cases
%\end{itemize} % alternatives 0
\end{pf}

\section{Splitting}\label{subsec:Split-lists}

\subsection{Split into minimum/rest of elements.}

The target functions are $\mn[X]$ which selects from $X$ the minimum element according
to the domain ordering and $\trim[X]$ which gives the list without it.
We need to prove \rcj{conj:min-Trim}.

\bpf{Min and Trim}{pf:list-min-trim}
By natural style proving, take $X_0$ arbitrary but fixed, assume:
\begin{equation}\label{assm:trim-list-1a}
  X_0\neq \el
\end{equation}
and after introducing the existential metavariables, the goal is:
\begin{equation}\label{goal:trim-list-0}
 \ms[X_0] = \ms[Y^*]\union \mse{y^*} \Aand y^*\leq X_0.
\end{equation}
Strategy \rst{st:cover-set} (cover set) applies to $X_0$, using only $a_0\cons U_0$
because (\ref{assm:trim-list-1a}).
The goal is:
\begin{equation}\label{goal:trim-list-1}
 \ms[a_0\cons U_0] = \ms[Y^*]\union \mse{y^*} \Aand y^*\leq a_0\cons U_0.
\end{equation}
By \rir{ir:red-comp-arg} (composite argument) on the last conjunct the goal becomes:
\begin{equation}\label{goal:trim-list-2}
 \ms[a_0\cons U_0] = \ms[Y^*]\union \mse{y^*} \Aand y^*\leq a_0 \Aand y^*\leq U_0.
\end{equation}
Strategy \rst{st:cascading} (cascading) generates the conjecture:
\begin{conjecture}\label{conj:min-aux-trim-aux}
  $\fa{X}\fa{a}\ex{y}\ex{Y}(\ms[a\cons X] = \ms[Y]\union \mse{y}
  \Aand y\leq a \Aand y\leq X).$
\end{conjecture}
\rpf{pf:list-min-trim-aux} synthesizes the auxiliary functions $\mnaux$ and $\trimaux$ which have the property:
\begin{equation}\label{assm:min-aux-trim-aux}
  \fa{X}\fa{a}(\ms[a\cons X] = \ms[\trimaux[a,X]]\union \mse{\mnaux[a,X]}
  \Aand \mnaux[a,X]\leq a \Aand \mnaux[a,X]\leq X)
\end{equation}
and which solves the goal (\ref{goal:trim-list-2}) using the witnesses
$\{Y^*\rightarrow\trimaux[a_0,U_0],\ y^*\rightarrow\mnaux[a_0,U_0]\}.$
\end{pf}

\smallskip
\noindent
We prove now \rcj{conj:min-aux-trim-aux}.

\bpf{Min and Trim auxiliary}{pf:list-min-trim-aux}
By quantified inferences the goal becomes:
\begin{equation}\label{goal:min-aux-trim-aux-1}
 \ms[a_0\cons X_0] = \ms[Y^*]\union \mse{y^*} \Aand y^*\leq a_0 \Aand y^*\leq X_0.
\end{equation}
Apply \rst{st:cover-set} (cover set) on $X_0.$
%\begin{itemize} % cases 1
  %\casitem{1}

  \noindent
  {\em Case 1.} $X_0= \el$ is straightforward, the solutions are: $\{y^*\rightarrow a_0,\ Y^*\rightarrow \el\}.$

  %\casitem{2}
  \noindent
  {\em Case 2.} $X_0 = b_0\cons U_0$ generates the goal:
  \begin{equation}\label{goal:min-aux-trim-aux-2}
    \ms[a_0\cons (b_0\cons U_0)] = \ms[Y^*]\union \mse{y^*}
    \Aand y^*\leq a_0 \Aand y^*\leq b_0\cons U_0.
  \end{equation}
  By \rir{ir:expand-multiset} (expand multiset) and \rir{ir:red-comp-arg}
  (reduce composite argument) the goal becomes:
  \begin{equation}\label{goal:min-aux-trim-aux-3}
    \mse{a_0} \union \mse{b_0} \union \ms[U_0] = \ms[Y^*]\union \mse{y^*}
    \Aand y^*\leq a_0 \Aand y^*\leq b_0 \Aand y^*\leq U_0.
  \end{equation}
  Two cases for domain element constants are generated by rule \rir{ir:two-constants}
  (two constants):
  \begin{itemize} % cases elementary constants
    \caseq{2.1}{a_0 \leq b_0}{assm:min-aux-trim-aux-0}
    Strategy \rst{st:induction} (induction) applies to $U_0, a_0$ in
    (\ref{goal:min-aux-trim-aux-1}) and add the assumption:
    \begin{equation}\label{assm:min-aux-trim-aux-1}
      \begin{split}
        \ms[U_0]\union \mse{a_0} = \ms[\trimaux[a_0,U_0]]\union \mse{\mnaux[a_0,U_0]}\ \ \Aand \\
        \mnaux[a_0,U_0]\leq a_0\ \  \Aand \mnaux[a_0,U_0]\leq U_0.
      \end{split}
    \end{equation}
    (\ref{goal:min-aux-trim-aux-3}) is rewritten by equality (\ref{assm:min-aux-trim-aux-1}):
    \begin{equation}\label{goal:min-aux-trim-aux-4}
      \begin{split}
        \ms[\trimaux[a_0,U_0]]\union \mse{\mnaux[a_0,U_0]} \union \mse{b_0}= \ms[Y^*]\union \mse{y^*}\ \  \Aand \\
        y^*\leq a_0\ \  \Aand\ \  y^*\leq b_0\ \  \Aand\ \  y^*\leq U_0.
      \end{split}
    \end{equation}
    The goal equation is split by strategy \rst{st:split-goal-eq}:
    \begin{equation}\label{goal:min-aux-trim-aux-4a}
      \begin{split}
        \ms[\trimaux[a_0,U_0]]\union \mse{b_0}= \ms[Y^*] \Aand \\
        \mse{\mnaux[a_0,U_0]} = \mse{y^*} \Aand y^*\leq a_0 \Aand y^*\leq b_0 \Aand y^*\leq U_0.
      \end{split}
    \end{equation}
    By \rir{ir:solve-meta} (solve metavariable) the solutions are:
    $\{y^*\rightarrow \mnaux[a_0,U_0],\ Y^*\rightarrow b_0\cons \trimaux[a_0,U_0]\}$ and the remaining goal is proven by standard logic and properties of ordering.

    \caseq{2.2}{b_0 < a_0}{assm:min-aux-trim-aux-0.2}
    The proof proceeds similarly by applying induction on $U_0, b_0$ in (\ref{goal:min-aux-trim-aux-1})) and the obtained solutions are: $\{y^*\rightarrow \mnaux[b_0,U_0],\ Y^*\rightarrow a_0\cons \trimaux[b_0,U_0]\}.$
  \end{itemize}
%\end{itemize}
%\end{itemize}
\end{pf}

The extracted algorithms from the proofs are:
\begin{myalgo}\label{algorithm:lists-minaux} {\em Minimum.}
  \mbox{ }\\
  \(\underset{a,b,U}{\forall }\left(
  \begin{array}{c}
    \mn[a\cons U] = \mnaux[a, U]\\
    \mnaux[a,\el]=a\\
    \mnaux[a,b\cons U]=
    \left\{\!\!\!
    \begin{array}{ll}
      \mnaux[a, U], & \mbox{if } a \leq b\\
      \mnaux[b, U], & \mbox{if } b < a\\
    \end{array}
    \right.
  \end{array}
  \right)\)
\end{myalgo}
\begin{myalgo}\label{algorithm:lists-trimaux} {\em Trim.}
  \mbox{ }\\
  \(\underset{a,b,U}{\forall }\left(
  \begin{array}{c}
    \trim[a\cons U] = \trimaux[a, U]\\
    \trimaux[a,\el]=\el\\
    \trimaux[a,b\cons U]=
    \left\{\!\!\!
    \begin{array}{ll}
      b\cons \trimaux[a, U], & \mbox{if } a \leq b\\
      a\cons \trimaux[b, U], & \mbox{if } b < a\\
    \end{array}
    \right.
  \end{array}
  \right)\)
\end{myalgo}

%\smallskip

%\noindent
\subsection{Split into smaller/bigger elements.}

We need functions $\smalleq[a,X]$ and $\bigger[a,X]$ which select from $X$ the
elements which are smaller or equal, respectively strictly bigger than $a$ according to
the domain ordering.
We prove \rcj{conj:lesseq-bigger}.
\bpf{Split}{pf:split}
  $a$ Skolemizes to $a_0$ and $X$ to $X_0$ (target constant), and the goal uses the metavariables $V^*,W^*$:
  \begin{equation}\label{goal:split-list-1}
    \ms[X_0] = \ms[V^*]\union\ms[W^*]\Aand
    V^* \leq a_0 \Aand a_0 < W^*.
  \end{equation}
  Strategy \rst{st:cover-set} applies to $X_0$ with cover set $\{\el,\ b_0\cons U_0\}$:
  %\begin{itemize}
    %\casitem{1}

    \noindent
    {\em Case 1.} $X_0 = \el$ is straightforward with solutions: $\{V^*\rightarrow \el,\ W^*\rightarrow \el\}.$

    %\casitem{2}

    \noindent
    {\em Case 2.} $X_0 = b_0\cons U_0$:
    \begin{equation}\label{goal:split-list-3}
      \ms[b_0\cons U_0] = \ms[V^*]\union\ms[W^*]\Aand
      V^* \leq a_0 \Aand a_0 < W^*.
    \end{equation}
    By \rir{ir:expand-multiset} (expand multiset):
    \begin{equation}\label{goal:split-list-4a}
      \mse{b_0}\union\ms[U_0] = \ms[V^*]\union\ms[W^*]\Aand
          V^* \leq a_0 \Aand a_0 < W^*.
    \end{equation}
    By \rst{st:induction} (induction) on $U_0$ (smaller than $X_0$) adds the assumption:
    \begin{equation}\label{assm:split-list-1}
      \ms[U_0] = \ms[\smalleq[a_0,U_0]]\union\ms[\bigger[a_0,U_0]]\Aand
      \smalleq[a_0,U_0] \leq a_0 \Aand a_0 < \bigger[a_0,U_0].
    \end{equation}
    By rewriting $\ms[U_0]$ in the goal:
     \begin{equation}\label{goal:split-list-4}
      \mse{b_0}\union\ms[\smalleq[a_0,U_0]]\union\ms[\bigger[a_0,U_0]]
      = \ms[V^*]\union\ms[W^*]\Aand
          V^* \leq a_0 \Aand a_0 < W^*.
    \end{equation}
     Inference rule \rir{ir:two-constants} (two constants) issues two cases:
     \begin{itemize} % cases elementary constants
      \caseq{2.1}{b_0 \leq a_0}{assm:a1-leq-a0}
      Strategy \rst{st:split-goal-eq} (split goal equation) changes the goal:
    \begin{equation}\label{goal:split-list-5}
      \mse{b_0}\union\ms[\smalleq[a_0,U_0]]
      = \ms[V^*]\Aand
          V^* \leq a_0,
    \end{equation}
    \begin{equation}\label{goal:split-list-6}
      \ms[\bigger[a_0,U_0]]
      = \ms[W^*]\Aand
          a_0 < W^*.
    \end{equation}
    By \rir{ir:compress-multiset} (compress multiset) in (\ref{goal:split-list-5}), and by \rir{ir:solve-meta} (solve metavariable) in both (\ref{goal:split-list-5}) and (\ref{goal:split-list-6}), the obtained solutions are:
     $\{V^*\rightarrow b_0\cons \smalleq[a_0,U_0],\ W^*\rightarrow \bigger[a_0,U_0]\}$ and the remaining goal is proven by standard inferences.

      \caseq{2.2}{a_0 < b_0}{assm:a0-less-a1}
      Similarly, the obtained solutions are: $\{V^*\rightarrow\smalleq[a_0,U_0],\ W^*\rightarrow b_0\cons \bigger[a_0,U_0]\}.$
    \end{itemize} % cases elementary constants
  %\end{itemize} % cases cover-set
\end{pf}

%The extracted algorithms from the proof are:
\begin{myalgo}\label{algorithm:lists-smalleq} {\em Small or equal}
\mbox{ }\\
\(\underset{a,b,U}{\forall }\left(
\begin{array}{c}
\smalleq[a,\el]=\el\\
\smalleq[a,b\cons U]=
\left\{\!\!\!
              \begin{array}{ll}
              b\cons \smalleq[a, U], & \mbox{if } b \leq a\\
              \smalleq[a, U], & \mbox{if } a < b\\
              \end{array}
\right.
\end{array}
\right)\)
\end{myalgo}

\begin{myalgo}\label{algorithm:lists-bigger} {\em Bigger}
\mbox{ }\\
\(\underset{a,b,U}{\forall }\left(
\begin{array}{c}
\bigger[a,\el]=\el\\
\bigger[a,b\cons U]=
\left\{\!\!\!
              \begin{array}{ll}
              \bigger[a, U], & \mbox{if } b \leq a\\
              b\cons \bigger[a, U], & \mbox{if } a < b\\
              \end{array}
\right.
\end{array}
\right)\)
\end{myalgo}

\section{Merging}\label{sec:merge}

For lack of space we cannot present here the synthesis proofs for \ins\ and \conc,
the generated algorithms are the standard well known recursive ones.
We focus instead on the merging of two sorted lists into a sorted one, which is more interesting because many alternative algorithms are produced.

\bpf{Merge}{pf:merge} The goal \rcj{conj:merge} is Skolemized ($X_0$ is the target constant),
and the target goal is:
\begin{equation}\label{goal:merge-g0}
\fa{Y}((\is[X_0] \And \is[Y]) \Implies \ex{W}(\ms[W] = \ms[X_0]\union\ms[Y] \Aand \is[W])).
\end{equation}
After Skolemizing $Y$ to $Y_0$ the LHS of the implication becomes assumption, and the RHS becomes goal and uses the metavariable $W^*$:
\begin{equation}\label{goal:merge-g1}
\ms[W^*] = \ms[X_0]\union\ms[Y_0] \Aand \is[W^*].
\end{equation}
By \rst{st:cover-set} (cover set) on $X_0:$
%\begin{itemize} % cases 1
%\casitem{1}
\\ \noindent {\em Case 1:} $X_0 = \el$. By straightfoward proof the solution is $\{ W^* \rightarrow Y_0\}.$
%\casitem{2}
\\ \noindent {\em Case 2:} $X_0 = a_0\cons U_0$.
\\ \noindent By \rir{ir:expand-multiset} (expand multiset) on $\ms[a_0\cons U_0]$ the goal becomes:
\begin{equation}\label{goal:merge-g9}
  \ms[W^*] = \mse{a_0}\union \ms[U_0]\union\ms[Y_0] \Aand \is[W^*].
\end{equation}
\begin{itemize}
\altitem{2.1}
By strategy \rst{st:induction} (induction) which uses $U_0$ (smaller than $X_0$):
\begin{equation}\label{assm:merge-a10.1}
  \ms[\merge[U_0, Y_0]] = \ms[U_0]\union \ms[Y_0]\Aand
  \is[\merge[U_0,Y_0]].
\end{equation}
By rewriting using (\ref{assm:merge-a10.1}) the goal (\ref{goal:merge-g9}) becomes:
\begin{equation}\label{goal:merge-g11}
  \ms[W^*] = \mse{a_0}\union \ms[\merge[U_0,Y_0]] \Aand \is[W^*].
\end{equation}
Application of \rst{st:pair-MS} (pair multisets) on $\mse{a_0}$ and $\ms[\merge[U_0,Y_0]]$ and of \rst{st:cascading} (cascading) using (\ref{assm:merge-a10.1}) and (\ref{goal:merge-g11}) produces \rcj{conj:insert} which is used to generate the algorithm \ins:
\begin{equation}\label{goal:merge-g14}
  \ms[W^*] = \ms[\ins[a,\merge[U_0,Y_0]]] \Aand \is[W^*].
\end{equation}
By \rir{ir:solve-meta} the solution is $\{ W^* \rightarrow \ins[a_0, \merge[U_0,Y_0]]\}$
and the synthesized algorithm is:
\begin{myalgo}\label{algorithm:sorted-lists-merge-1} {\em Merge sorted lists using insert, version 1.}
\mbox{ }\\
\(\underset{a,U,V}{\forall }\left(
\begin{array}{c}
\merge[\el, V]=V\\
\merge[a\cons U, V]= \ins[a, \merge[U,V]]
\end{array}
\right)\)
\end{myalgo}

This is of course not the most efficient algorithm because the induction is not used
on both arguments (as it is done in the sequel, see below).
A hint about inefficiency is that the property of $U$ to be sorted is not used in the
proof, but this has also a positive side:
$Merge[U, \el]$ is a sorting algorithm, essentially equivalent to {\em insert sort}.

\altitem{2.2} Applying strategy \rst{st:pair-MS} (pair multisets) to $\mse{a}$ and $\ms[V_0]$ and then \rst{st:cascading} (cascading) produces the same conjecture for \ins,
and the goal becomes:
\begin{equation}\label{ec:concat-lists-sorted-g2-cascade-v2}
\ms[W^*] = \ms[U_0]\union \ms[\ins[a, V_0]] \Aand \is[W^*]
\end{equation}
with the additional assumption: $\is[Ins[a,V_0]].$
We can apply now strategy \rst{st:induction} (induction) to the pair of multiset
terms and construct the list $U_1$ which is sorted and whose multiset is equal to the union.
Therefore the solution is $\{W^* \rightarrow U_1\}$
and the merging algorithm is:
\begin{myalgo}\label{algorithm:sorted-lists-merge-2} {\em Merge sorted lists using insert, version 2.}
\mbox{ }\\
\(\underset{a,U,V}{\forall }\left(
\begin{array}{c}
\merge[\el, V]=V\\
\merge[a\cons U, V]= \merge[U,\ins[a,V]]
\end{array}
\right)\)
\end{myalgo}
This algorithm, although not optimal, is interesting because it is tail--recursive,
and, since only the second argument needs to be sorted, it can also be used for sorting
as $\merge[U, \el],$ which is again {\em insert sort}.

\textbf{Remark.} If the proof continues from the goal (\ref{goal:merge-g9}) by applying strategies
\rst{st:pair-MS} (pair multisets) to $\mse{a}$ and $\ms[U_0],$and then \rst{st:cascading} (cascading), then induction cannot be applied to the resulting multiset pair ($\ms[\ins[a_0,U_0]]$ and $\ms[Y_0]$) because $\ins[a_0,U_0]$ is not smaller than the target constant $X_0 = a_0 \cons U_0$. The corresponding algorithm would have $\merge[a\cons U, V]= \merge[\ins[a,U],V]$ as the second clause, which is an infinite loop.
\altitem{2.3} The proof continues from goal (\ref{goal:merge-g9}) by applying \rst{st:cover-set} (cover set) on $Y_0$ in a {\em nested} fashion: now we have a second target constant $Y_0$ and a second target goal obtained from (\ref{goal:merge-g0}):
\begin{equation}\label{goal:merge-y0}
(\is[a_0\cons X_0] \And \is[Y_0]) \Implies \ex{W}(\ms[W] = \ms[a_0 \cons X_0]\union\ms[Y_0] \Aand \is[W]).
\end{equation}
This is not a goal in the proof, but a pattern for generating new assumptions by induction, for ground terms smaller than $Y_0$, by strategy \rst{st:induction}.
\begin{itemize} % cases 1
\casitem{2.3.1} $Y_0 = \el$:
Similarly, the solution is $\{ W^* \rightarrow a_0\cons U_0\}.$
\casitem{2.3.2} $Y_0 = b_0\cons V_0$:
By application of \rir{ir:red-comp-arg} (reduce composite argument) to $\is[Y_0]$:
\begin{equation}\label{assm:merge-a39}
\is[b_0\cons V_0] \Aand b_0\leq V_0 \Aand \is[V_0]
\end{equation}
and the goal becomes:
\begin{equation}\label{goal:merge-g42}
  \ms[W^*] = \mse{a_0}\union \ms[U_0]\union \mse{b_0}\union \ms[V_0] \Aand \is[W^*].
\end{equation}
When the rule \rir{ir:two-constants} (two constants) is applied, then one has:
\begin{itemize}
	\casitem{2.3.2.1} $a_0\leq b_0$: A successful proof alternative proceeds by using first \rir{ir:compress-multiset} to replace $\mse{b_0}\union \ms[V_0]$ by $\ms[b_0 \cons V_0]$,
	then by using \rst{st:pair-MS} (pair multisets) and \rst{st:induction} (induction) on (\ref{goal:merge-g0}) to replace $\ms[U_0]\union\ms[b_0 \cons V_0]$ by $\ms[\merge[U_0,b_0 \cons V_0]]$ (because $U_0$ is less than $X_0$, and as second argument of \merge\ any $Y$ is allowed).
	After that, prefixing $a_0$ to this by \rir{ir:compress-multiset} (compress multiset) results in a sorted list
	by the current assumptions and the properties of the domain ordering.
	\casitem{2.3.2.2} $b_0 < a_0$: A similar proof alternative succeeds, but here
	\rst{st:pair-MS} (pair multisets) is applied to $\ms[a_0 \cons U_0]$ and $V_0$ and then \rst{st:induction} (induction) can be used on the basis of the second target goal (\ref{goal:merge-y0}), because the second argument $V_0$ is less than $Y_0$ and the first argument is exactly	as in the pattern. Finally the algorithm is the classical one:
\end{itemize}
\begin{myalgo}\label{algorithm:sorted-lists-merge-3} {\em Merge sorted lists, version 3.}
\mbox{ }\\
\(\underset{a,b,U,V}{\forall }\left(
\begin{array}{c}
\merge[\el, V]=V\\
\merge[a\cons U, \el] = a\cons U\\
\merge[a\cons U, b\cons V]= \left\{
              \begin{array}{ll}
              a\cons \merge[U, b\cons V], & \mbox{if } a \leq b\\
                b\cons \merge[a\cons U, V], & \mbox{if } b < a\\
              \end{array}
\right.
\end{array}
\right)\)
\end{myalgo}
Because there are 4 multiset terms in the goal, strategy \rst{st:pair-MS} (pair multisets) generates many alternatives, which in turn lead to several algorithms, which only differ in the RHS of the last clause, but are less efficient than the one above.
Some of them have interesting properties, for instance the one ending in
$\merge[a\cons U, b\cons V]= \ins[a,\ins[b,\merge[U,V]]]$
will generate a sorted list even if the arguments are not sorted,
while the one ending in
$\merge[a\cons U, b\cons V]= \ins[b,\merge[a\cons U, V]]$
needs only the first argument to be sorted.
\end{itemize}
\end{itemize}
%\end{itemize}
\end{pf}

\section{Conclusions and Further Work}\label{sec:conclusions}

We demonstrate the possibility of automatic synthesis of complex
algorithms on (possibly sorted) lists, using the notion of multiset.
The proofs are more efficient than by using general resolution,
because specific inference rules and strategies which are also taylored
for synthesis proofs, notably for discovering concrete induction principles
and for synthesizing needed auxiliary functions.
The various algorithms which are produced can constitute a test field for
methods of automatic evaluation of efficiency, time and space consumption, etc.
A distinctive feature of our approach is the use of natural--style proofs, which is
supported by the \tma system.
The natural style of proving (as formula notation, as proof text, and as inference steps)
has the advantage of allowing human inspection in an intuitive way, and this
facilitates the development of intuitive inference rules which embed the knowledge
about the underlying domains.
The experiments presented here continue our previous work on synthesis of deletion algorithms,
as well as merging and inserting on lists and trees, and are a prerequisite for further work on
synthesis of more complex algorithms for sorting and searching, including  operations on several domains.

\bibliographystyle{plainurl}
\balance
\bibliography{bibliografie}

\end{document}